\documentclass[twocolumn,preprintnumbers,amsmath,amssymb]{revtex4}
\usepackage{amsmath, amsthm, amssymb,graphicx,mathtools,algpseudocode,algorithmicx,color}
\begin{document}
\title{Fast simulation of Brownian dynamics in a crowded environment}
\author{Stephen Smith}
\affiliation{School of Biological Sciences, University of Edinburgh, Mayfield Road, Edinburgh EH9 3JR, Scotland, UK}

\author{Ramon Grima}
\affiliation{School of Biological Sciences, University of Edinburgh, Mayfield Road, Edinburgh EH9 3JR, Scotland, UK}
\begin{abstract}
Brownian dynamics simulations are an increasingly popular tool for understanding spatially-distributed biochemical reaction systems. Recent improvements in our understanding of the cellular environment show that volume exclusion effects are fundamental to reaction networks inside cells. These systems are frequently studied by incorporating inert hard spheres (crowders) into three-dimensional Brownian dynamics simulations, however these methods are extremely slow owing to the sheer number of possible collisions between particles. Here we propose a rigorous ``crowder-free'' method to dramatically increase simulation speed for crowded biochemical reaction systems by eliminating the need to explicitly simulate the crowders. We consider both the case where the reactive particles are point particles, and where they themselves occupy a volume. We use simulations of simple chemical reaction networks to confirm that our simplification is just as accurate as the original algorithm, and that it corresponds to a large speed increase.
\end{abstract}
\maketitle
\noindent

\section{Introduction}
The fact that living cells constitute crowded cytoplasmic and nuclear environments has been appreciated for several decades \cite{zimmerman1993macromolecular,ellis2001macromolecular}. However, the significance of excluded volume effects to specific biochemical processes has recently been highlighted by a multitude of experimental and theoretical observations. It is now established that crowding by large inert molecules can place limits on the total number of transcription factors in a cell \cite{li2009effects}, can cause DNA to change its shape \cite{zhang2009macromolecular}, can encourage protein structure self-assembly \cite{rivas2001direct}, and can both enhance and diminish transcription factor binding rates \cite{tan2013molecular}.

Correspondingly, several authors have recently proposed a variety of mathematical descriptions of crowding effects. Many of these are modifications of the compartment-based reaction-diffusion master equation \cite{cianci2015molecular,taylor2015reconciling,meinecke2016multiscale}, which divides space into a lattice and models diffusion as particles hopping between neighbouring lattice sites. Lattice-based models have, however, been shown to underestimate the effects of crowding compared to more detailed descriptions \cite{grima2006systematic,meinecke2016excluded}. Some authors have proposed introducing crowding effects directly into non-spatial descriptions such as the chemical master equation \cite{grima2010intrinsic} or the deterministic reaction rate equations \cite{berry2002monte,schnell2004reaction}. The most popular technique, however, involves Brownian dynamics (BD) simulations \cite{wieczorek2008influence,ando2010crowding,mcguffee2010diffusion}.

BD simulations explicitly track the positions of particles and model diffusion as a Brownian random walk in continuous space. Several popular modern BD simulators do not model crowding explicitly, since they assume particles to be point-particles with no physical volume \cite{andrews2004stochastic,van2005simulating}. However, designing algorithms to accurately study the behaviour of hard sphere colloids (uniform suspensions of insoluble particles) without hydrodynamic interactions was a popular problem in chemical physics long before the biochemical implications of volume exclusion were fully appreciated \cite{lowen1991brownian,schaertl1994brownian,cichocki1994friction}.

One such algorithm was proposed by Cichocki and Hinsen \cite{cichocki1990dynamic}. The idea behind the Cichocki-Hinsen algorithm is simple to state: only one particle is moved at a time, and if the attempted move results in a collision the particle is simply placed back in its previous position, thereby crudely modelling a steric repulsion. Despite its relative simplicity, the Cichocki-Hinsen algorithm has been proved to converge to the Smoluchowski equation in the limit of short simulation time-steps \cite{cichocki1990dynamic} and has been shown to agree perfectly with far more detailed algorithms which incorporate particle velocity and momentum \cite{strating1999brownian}. It is therefore commonly used to simulate Brownian diffusion of hard spheres \cite{doliwa1998cage,auer2001prediction,auer2004numerical}, yet because of its fine-grained detail, the algorithm must be run for a long time to get statistically significant results.

In this article, we propose a modification to the Cichocki-Hinsen algorithm for reaction-diffusion systems. Our simplification arises from distinguishing between reactive particles (which may either be point particles or have a finite volume) and hard sphere crowders. We rigorously derive the probability that a reactive particle will collide with a crowder in a single time step, and use this to write a modified Cichocki-Hinsen algorithm which does not explicitly simulate crowders: we call this the \emph{crowder-free} algorithm. Unsurprisingly, the crowder-free algorithm results in a dramatic speed increase over the original Cichocki-Hinsen algorithm of up to three orders of magnitude. Perhaps more surprisingly, the output data of the two algorithms is near-indistinguishable for each example that we test.

In section \ref{PointParticles} we propose the crowder-free algorithm for a system of reactive point particles in a sea of hard sphere crowders. We first outline the Cichocki-Hinsen algorithm for a point particle reaction-diffusion system. We then derive the probability that a small diffusive jump by a reactive point particle results in a collision with a crowder. Using this expression, we outline the crowder-free algorithm. We subsequently test our algorithm's speed and accuracy in modelling both pure diffusion and the reaction-diffusion system $A+B \xrightleftharpoons[]{}C$ in the presence of crowders. 

In section \ref{Finite} we analogously propose the crowder-free algorithm for a system of finite-size reactive particles in a sea of hard sphere crowders. We then derive the probability that a small diffusive jump by a finite-size reactive particle results in a collision with a crowder: this is shown to be very similar to the point particle expression. We again test our algorithm's speed and accuracy in modelling pure diffusion and the reaction-diffusion system $\emptyset \xrightarrow{} X,~ X+X \xrightarrow{} \emptyset$ in the presence of crowders. We conclude with a discussion in section \ref{Discussion}.
\section{Point particles in a crowded environment}\label{PointParticles}
We first describe the Cichocki-Hinsen algorithm as applied to a system of reactive point particles in a sea of inert spherical crowders of radius $R$. Since the original Cichocki-Hinsen algorithm was written for purely diffusive systems, we have added some steps for reactive systems. The reactive method we use is the Doi model \cite{doi1976stochastic,erban2009stochastic}, which assigns each bimolecular reaction $j$ a distance $r_j$ and a rate $\lambda_j$. Bimolecular reaction $j$ occur with rate $\lambda_j$ when two reactive particles of the relevant type come within a distance $r_j$ of each other. Unbinding reactions are assigned a rate $\lambda_j$ and an unbinding distance $\sigma_j$. These reactions occur with rate $\lambda_j$ and the daughter particles are placed a distance $\sigma_j$ from each other, at a uniformly distributed angle. Other monomolecular and zero-order reactions are simply assigned a rate $\lambda_j$. Note that reaction distances and unbinding distances are not physical radii, and do not exclude any volume.\\~\\
\textbf{Cichocki-Hinsen algorithm with reactive point particles}
\begin{enumerate}
\item Uniformly distribute the reactive particles and the crowders in the volume, such that no crowders are intersecting each other and no reactive particles lie inside a crowder. Let $N$ be the total number of particles (reactive and crowders), and randomly assign each particle a unique index $1,...,N$.

\item For each $i=1,...,N$, propose a new position for particle $i$ at a random Normal$(0,\sqrt{2 D_i \Delta t})$ displacement in each spatial dimension, where $D_i$ is the diffusion coefficient of particle $i$ and $\Delta t$ is the simulation time step. If this new position causes an intersection between any particles (reactive and crowder), place particle $i$ back in its original position. If not, place particle $i$ in the new position.

\item For each reactive particle involved in a bimolecular reaction $j$, check if any reactive particles of the appropriate types lie inside a sphere of radius $r_j$ around the particle. For each appropriate reactive particle inside this sphere, propose a reaction with probability $\lambda_j \Delta t$. If successful, check if any daughter particles would intersect a crowder. If so, skip the reaction; if not, allow the reaction to proceed.

\item For each reactive particle of a type involved in a unimolecular reaction $j$, propose a reaction with probability $\lambda_j \Delta t$. If successful, check if any daughter particles would intersect a crowder. If so, skip the reaction; if not, allow the reaction to proceed.

\item For each zero-order reaction, propose a reaction with probability $\lambda_j \Delta t$. If successful, check if any of the new particles would intersect a crowder. If so, skip the reaction; if not, allow the reaction to proceed.

\item Advance time by $\Delta t$. Let $N$ be the new total number of particles and randomly reassign each particle a unique index $1,...,N$. Return to (2) and repeat until a target time has elapsed.
\end{enumerate}

The overwhelmingly time-consuming step of this algorithm is step (2), in which potential particle overlaps must be checked $N$ times. The reaction steps (3)-(5) also involve potential overlaps, but as $\Delta t$ should typically be taken small enough that at most one reaction could plausibly happen per time step, these should not be particularly time-consuming. Our aim in the next subsection is therefore to reduce the time taken by step (2). Note that step (1) can also be particularly time-consuming: although our simplification does not particularly aim to fix that problem, it happens that by increasing the speed of step (2) we also dramatically shorten step (1).

\subsection{Derivation}
We first make two observations which form the basis of our method of reducing the time taken by the Cichocki-Hinsen algorithm. Firstly, the crowders are inert and contribute little to the actual reactive behaviour of the system; their only function is to occasionally prevent a reactive particle from moving or reaction from happening. Secondly, the crowders are uniformly distributed in space: this implies that each proposed reactive particle movement has roughly the same chance of being impeded by a crowder. 

One common method of modelling diffusion in a crowded environment, based on the crowder uniformity assumption, is to simply replace the diffusion coefficient $D$ with $D(1-\phi)$, where $\phi$ is the proportion of the total volume occupied by crowders \cite{fanelli2010diffusion}. The idea is that if a particle attempts to move to a new location, there is a $1-\phi$ probability of that location not being occupied by a crowder. This is a valid assumption if the particle displacement at a time step $\delta x \gg R$, that is, if the particle moves by a distance much greater than the crowder size, such that its new location can be roughly considered a uniform random variable. However, it makes little sense to take $\delta x \gg R$, because that would allow particles to pass through crowders with a single jump. 

On the other hand, taking $\delta x \ll R$ makes physical sense, because the tiny perturbations which make up Brownian motion are much smaller than any particle radius. Furthermore, this is precisely the limit in which Cichocki and Hinsen proved their algorithm to be exact \cite{cichocki1990dynamic}. In that limit, however, we cannot use the $1-\phi$ assumption. To understand why not, consider that the particle is already in a permitted location: this implies that there is a small sphere with radius $\epsilon>0$ around the particle which does not intersect any crowders. This local effect implies that the particle's new position cannot be treated as uniformly distributed: if $\delta x$ is small enough ($\delta x<\epsilon$), the particle's new position is guaranteed to not intersect any crowders. In summary, if we require that $\delta x \ll R$, then the probability that the particle's new position is illegal (intersects a crowder) is not given by $1-\phi$ but by some function of $\delta x$. We now attempt to derive that function.
\begin{figure}[h]
\centering
 \includegraphics[trim=0cm 0cm 4cm 0cm, clip=true,scale=0.348]{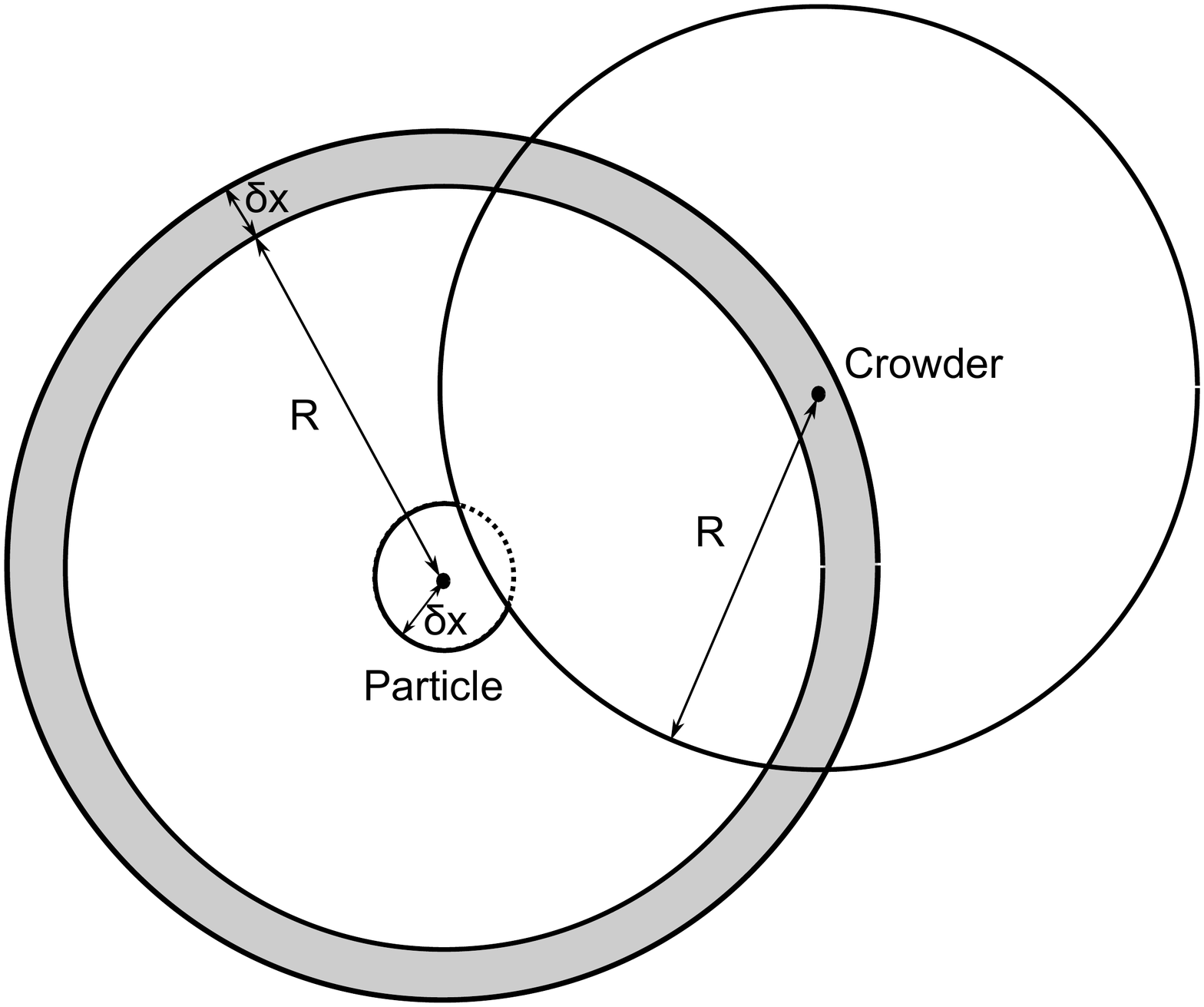}
 \caption{Diagram of a point particle attempting to move near a crowder of radius $R$. The particle attempts to displace itself a distance $\delta x$, such that its future position is on the surface of sphere of radius $\delta x$ around its current position. There may be crowders with their centres in the spherical shell of radius $R+\delta x$ (grey region), which could prevent the particle displacement. The proposed position will be illegal if it is on the dotted segment of the sphere of radius $\delta x$.}\label{diag1}
 \end{figure}
 
Consider what happens when a point-particle proposes to move by a displacement $\delta x$. This is illustrated in Fig. \ref{diag1}. The particle's proposed new position will be on the surface of a sphere of radius $\delta x$ around its current position. There will be no crowders with their centres in a sphere of radius $R$ around the particle (otherwise the point particle could not be where it is currently), however there is a non-zero probability that there are crowders with their centres inside the spherical shell between the sphere of radius $R+\delta x$ and the sphere of radius $R$ (the grey region in Fig. \ref{diag1}). If there are crowders in this region, then there is some probability that the point particle's proposed new position is illegal: this is precisely the probability that the proposed position intersects the crowder (the dotted line segment in Fig. \ref{diag1}).

\begin{figure}[h]
\centering
 \includegraphics[trim=1.2cm 13.8cm 2cm 6.3cm, clip=true,scale=0.49]{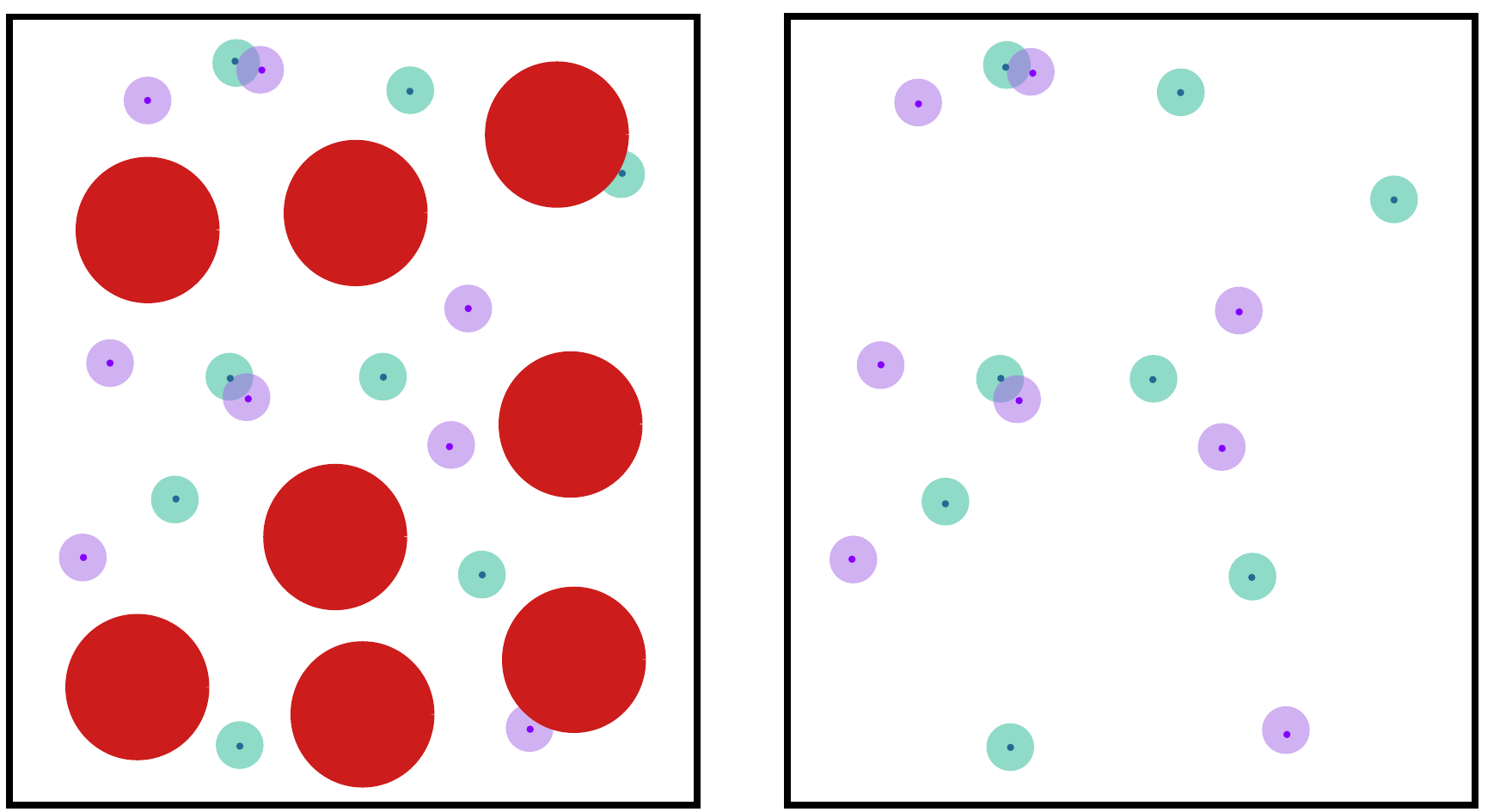}
 \caption{Cartoons of the Cichocki-Hinsen algorithm (left) and the crowder-free algorithm (right) for reactive point particles. The point particles (blue, purple) may have a reaction radius (translucent circle) which does not exclude any volume and is therefore permitted to intersect crowders (red) or other particles. The centres of the point particles (solid dots) are not permitted to intersect crowders.}\label{diag2}
 \end{figure}

Now, suppose that there are $N_C$ crowders of radius $R$ inside a volume $V$. Assuming a uniform crowder distribution, the probability that a given crowder is at risk of intersecting the point particle is simply the ratio of the volume of the grey region to the total volume:
\begin{equation}
p=\frac{\frac{4}{3} \pi (R+\delta x)^3-\frac{4}{3} \pi R^3}{V}=\frac{4\pi R^2 \delta x}{V}+o\left(\frac{\delta x}{R}\right).
\end{equation}
The probability of finding $n$ crowders in the grey region is then given by the Binomial distribution:
\begin{equation}\label{Pncrowders}
P(n\text{ crowders})= \frac{N_C!}{n!(N_C-n)!} p^n(1-p)^{N_C-n}.
\end{equation}
Of course, Eq. \eqref{Pncrowders} is only valid for small $n$, because there is a physical limit to how many crowders can fit in the relevant region. However, this is of little concern, since we are only concerned with the probabilities up to $o\left(\frac{\delta x}{R}\right)$, which turns out to correspond only to $n=0$ and $n=1$.
\begin{align}\label{P1crowder}
P(0\text{ crowders})&=1-\frac{4\pi N_CR^2 \delta x}{V}+o\left(\frac{\delta x}{R}\right),\\
P(1\text{ crowder})&=\frac{4\pi N_C R^2 \delta x}{V}+o\left(\frac{\delta x}{R}\right).\label{P2crowder}
\end{align}
We now consider the probability that the proposed new point particle position intersects the crowder. This is given by the surface area of the spherical cap of the sphere of radius $\delta x$ which lies inside the sphere of radius $R$ around the crowder (the dotted line segment in Fig. \ref{diag1}) divided by the total surface area of the sphere of radius $\delta x$. This is given by:
\begin{equation}\label{Pintersect}
P(\text{intersect})=\frac{2 \pi \delta x \frac{(R-\delta x+ d)(R+\delta x - d)}{2d}}{4 \pi \delta x^2},
\end{equation}
where $d$ is the separation between the centres of the point particle and the crowder \cite{mathworld}. The expected value of $d$ is simply $R+\frac{\delta x}{2}$, so inserting this into Eq. \eqref{Pintersect} gives:
\begin{equation}\label{Pintersect2}
P(\text{intersect})=\frac{1}{4}-\frac{3 \delta x}{16 R}+o\left(\frac{\delta x}{R}\right).
\end{equation}
Combining Eq. \eqref{P2crowder} with Eq. \eqref{Pintersect2} gives the probability that the proposed move is illegal:
\begin{equation}\label{pille}
P(\text{illegal})=\frac{4\pi N_C R^2 \delta x}{V}\left(\frac{1}{4}-\frac{3 \delta x}{16 R}\right)=\frac{\pi N_C R^2 \delta x}{V}+o\left(\frac{\delta x}{R}\right).
\end{equation}
Writing this in terms of the proportion of occupied volume, $\phi=\frac{\frac{4}{3} \pi N_C R^3}{V}$, leads to the simplified expression:
\begin{equation}\label{pillegal}
P(\text{illegal})=\frac{3\phi \delta x}{4R}+o\left(\frac{\delta x}{R}\right).
\end{equation}
We can therefore write a much faster version of Cichocki-Hinsen algorithm which \emph{does not include any crowders}. Only point particles need to be modelled explicitly in our algorithm, while the effect of crowders is incorporated by denying a point particle's proposed movement with probability $P(\text{illegal})$. For obvious reasons, we call this a \emph{crowder-free} algorithm. This idea is shown in Fig. \ref{diag2}. The left panel shows the Cichocki-Hinsen algorithm with crowders (red) and point particles (blue, purple). The points are not allowed to intersect the crowders, but the reaction radii are. The right panel shows the crowder-free algorithm, which looks identical to Cichocki-Hinsen without crowders. It is clear that the crowder-free algorithm will be easier to simulate.

Since none of the remaining particles in the crowder-free algorithm occupy any volume, we can move all particles simultaneously. The algorithm therefore essentially reduces to the classical Doi algorithm, with an extra clause for preventing particle movement. 
Some minor changes must also be made to the reaction parts of the algorithm (steps (3)-(5)), which originally prevented a reaction if a newly created particle would intersect a crowder. Since we no longer explicitly model crowders, we must modify this step. If the reaction is either bimolecular or monomolecular, the new particle will be placed at a small displacement $\sigma$ from a previous particle location. If $\frac{\sigma}{R} \ll 1$, then we can simply modify the diffusion formula to become $P(\text{illegal})=\frac{3\phi \sigma}{4R}$. Can we assume that $\frac{\sigma}{R} \ll 1$? In some cases, such as monomolecular conversion reaction of the type $A \rightarrow B$, we will have $\sigma = 0$, and it would be absurd to prevent such reactions due to crowding. However, some reactions may have quite a large unbinding distance, and the diffusion formula may prove to be invalid. At each such reaction, we therefore check if $\frac{\sigma}{R} < 0.1$. If this condition is true, we use the formula $P(\text{illegal})=\frac{3\phi \sigma}{4R}$, otherwise we use the formula $P(\text{illegal})=\frac{\frac{4}{3}\pi N_CR^3}{V}$, which is the probability that a uniformly distributed point particle would intersect a crowder. The choice of $0.1$ is essentially arbitrary, and can obviously be made smaller if required; we find that it gives good results, however. For zero-order reactions, we always use the formula $P(\text{illegal})=\frac{\frac{4}{3}\pi N_CR^3}{V}$, since particles created by these reactions have no parent particles.\\~\\
\textbf{Crowder-free algorithm with reactive point particles}
\begin{enumerate}

\item Uniformly distribute the reactive particles in the volume. 

\item Propose new positions for all particles at a random Normal$(0,\sqrt{2 D_i \Delta t})$ displacement in each spatial dimension, where $D_i$ is the diffusion coefficient of particle $i$ and $\Delta t$ is the simulation time step. Calculate $\delta x$, the length of the displacement, for each particle. With probability $\frac{3\phi \delta x}{4R}$ reject the proposed move, otherwise accept it.

\item For each particle of a type involved in a bimolecular reaction $j$, check if any particles of the appropriate types lie inside a sphere of radius $r$ around the particle, where $r$ is the reaction radius for the relevant reaction. For each appropriate particle inside this sphere, propose the reaction with probability $\lambda_j \Delta t$, where $\lambda_j$ is the corresponding reaction rate. For each daughter particle, calculate $\sigma$, the length of the displacement from the nearest parent particle. If $\frac{\sigma}{R}<0.1$, with probability $\frac{3\phi \sigma}{4R}$ reject the proposed reaction, otherwise accept it. Otherwise if $\frac{\sigma}{R} \geq 0.1$, with probability $\frac{\frac{4}{3}\pi N_CR^3}{V}$ (where $N_C$ is the number of crowders) reject the proposed reaction, otherwise accept it.

\item For each reactive particle of a type involved in a unimolecular reaction, propose a reaction with probability $\lambda_j \Delta t$, where $\lambda_j$ is the reaction rate. For each daughter particle, calculate $\sigma$, the length of the displacement from the nearest parent particle.  If $\frac{\sigma}{R}<0.1$, with probability $\frac{3\phi \sigma}{4R}$ reject the proposed reaction, otherwise accept it. Otherwise if $\frac{\sigma}{R} \geq 0.1$, with probability $\frac{\frac{4}{3}\pi N_C R^3}{V}$ reject the proposed reaction, otherwise accept it.

\item For each zero-order reaction, propose a reaction with probability $\lambda_j \Delta t$, where $\lambda_j$ is the reaction rate. With probability $\frac{\frac{4}{3}\pi N_CR^3}{V}$ reject the proposed reaction, otherwise accept it.

\item Advance time by $\Delta t$. Return to (2) and repeat until a target time has elapsed.
\end{enumerate}

In the next section, we confirm that the crowder-free algorithm is orders of magnitude faster than Cichocki-Hinsen, while retaining its accuracy.
\subsection{Comparative tests}\label{pptest}
In our first test of the crowder-free algorithm, we consider a single point particle diffusing in space, surrounded by a uniform distribution of crowders. This is the scenario for which the crowder-free algorithm should show the most dramatic improvement over the original Cichocki-Hinsen algorithm in terms of computation time.
\begin{figure}[h]
\includegraphics[scale=0.3]{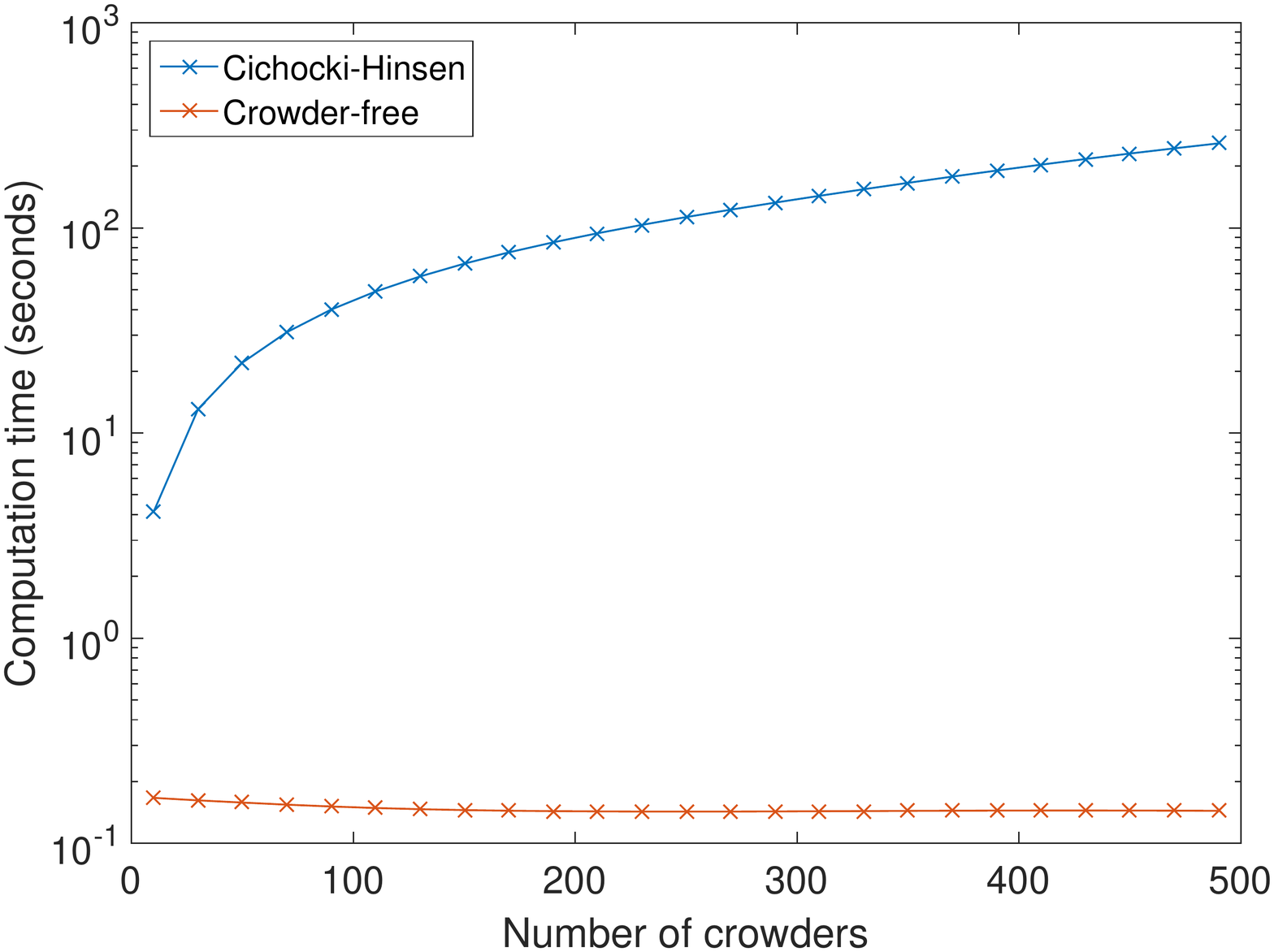}
\caption{Time taken for 100 time steps of both the Cichocki-Hinsen algorithm (blue) and the crowder-free algorithm (red), for a single point particle diffusing in space. With only 10 crowders, the crowder-free algorithm is over 10 times faster. With 500 crowders, the crowder-free algorithm is over $10^3$ times faster. Parameter values are $V=1$, $R=0.05$, $\Delta t=10^{-5}$, $D=0.1$ for the point particle, $D=0.01$ for the crowders.}\label{fig1}
\end{figure}
 Indeed, as shown in Fig. \ref{fig1}, we find that the crowder-free algorithm is at least an order of magnitude faster than the standard algorithm when there are only 10 crowders, this increases to three orders of maginitude when there are 500 crowders. A significant advantage is that the crowder-free algorithm does not scale with number of crowders, making it particularly useful for studying high levels of crowding.

Of course, fast simulation is of little use if the results of the algorithm are inaccurate. In our second test, we therefore use sample paths from both algorithms to compute the effective short-time diffusion coefficient $D^*$ of a single point particle in crowded space \cite{saxton1994anomalous}. This is done by performing a simulation with input diffusion coefficient $D$, computing the squared displacement of the particle at each time step and taking the mean of that value over the entire simulation. This value is equated to $6 D^* \Delta t$ to find an estimate for the effective short-time diffusion coefficient $D^*$. 
\begin{figure}[h]
\includegraphics[scale=0.3]{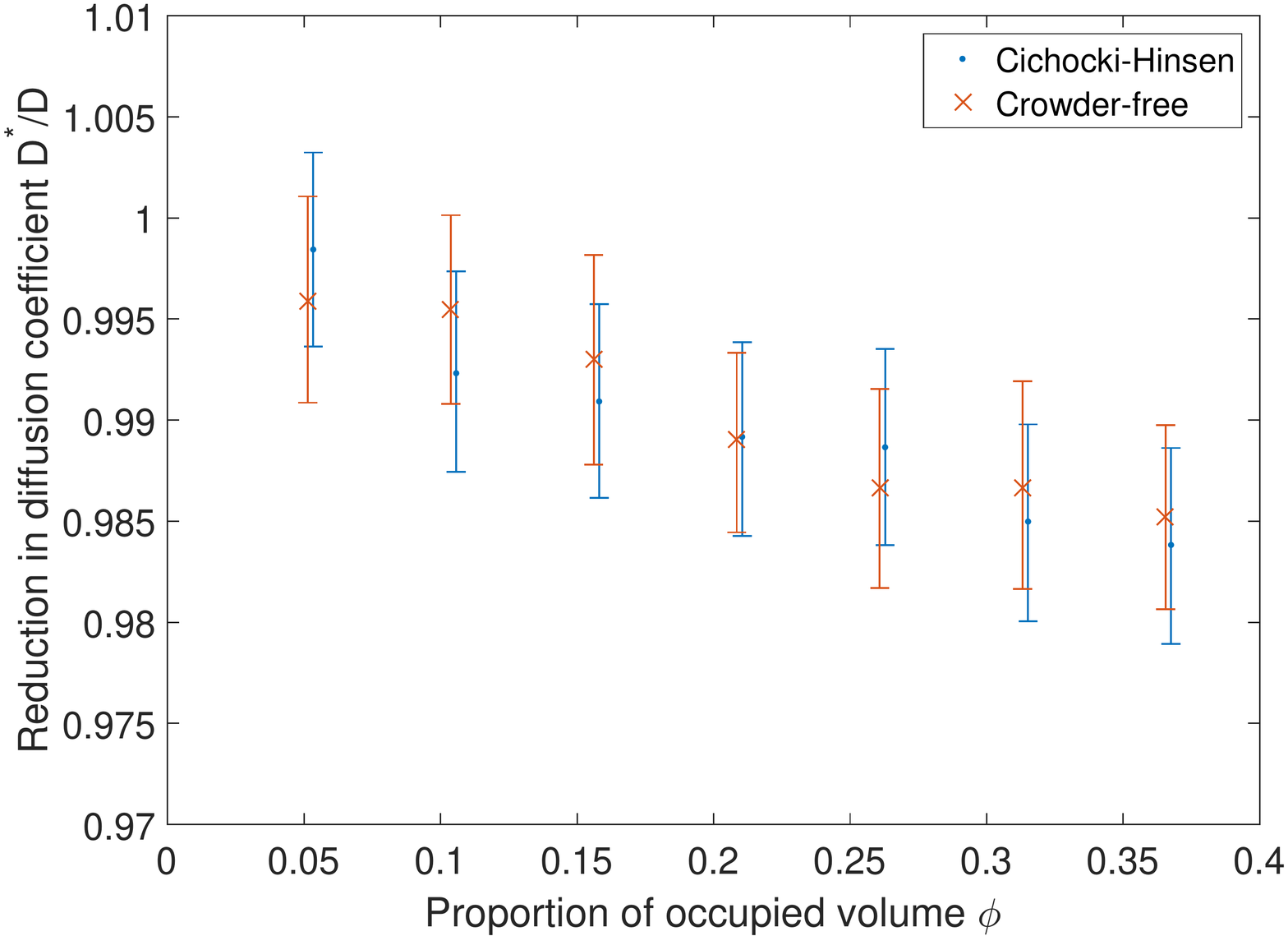}
\caption{Relative reduction in short-time diffusion coefficient for both the Cichocki-Hinsen algorithm (blue) and the crowder-free algorithm (red), for a single point particle diffusing in space, as a function of the proportion of occupied volume $\phi$. All data points are an average of 10 simulations, error bars are 1 standard deviation. Parameter values are $V=1$, $R=0.05$, $\Delta t=10^{-5}$, $D=0.1$ for the point particle, $D=0.01$ for the crowders.}\label{fig2}
\end{figure}
The non-dimensional parameter $\frac{D^*}{D}$ is the effective reduction in short-time diffusion coefficient due to crowding. For no crowding, we expect $\frac{D^*}{D}=1$, and the value should decrease as crowding increases. This is because large jumps are more likely to result in a collision with a crowder than small jumps, so the effective diffusion coefficient appears to be reduced. In Fig. \ref{fig2} we plot $\frac{D^*}{D}$ as a function of the proportion of occupied volume $\phi$. As expected, both algorithms show a reduction in the effective short-time diffusion coefficient as crowding increases, and both algorithms give very similar results, with their error bars always intersecting. Each data point is an average of 10 simulations, each simulation ran until the point particle, initially located at $(\frac{1}{2},\frac{1}{2},\frac{1}{2})$, left the unit cube with corners at $(0,0,0)$ and $(1,1,1)$.

We have confirmed that the crowder-free algorithm simulates diffusion as accurately as the original Cichocki-Hinsen algorithm, but we have not tested whether it accurately simulates reactions. In our next test, we use both algorithms to compute the equilibrium distribution of the reaction $A+B \xrightleftharpoons[]{}C$ in the presence of low and high levels of crowding. We expect the typical number of $C$ molecules to be higher for high crowding, because the unbinding reaction will occur less frequently. 
\begin{figure}[h]
\includegraphics[scale=0.3]{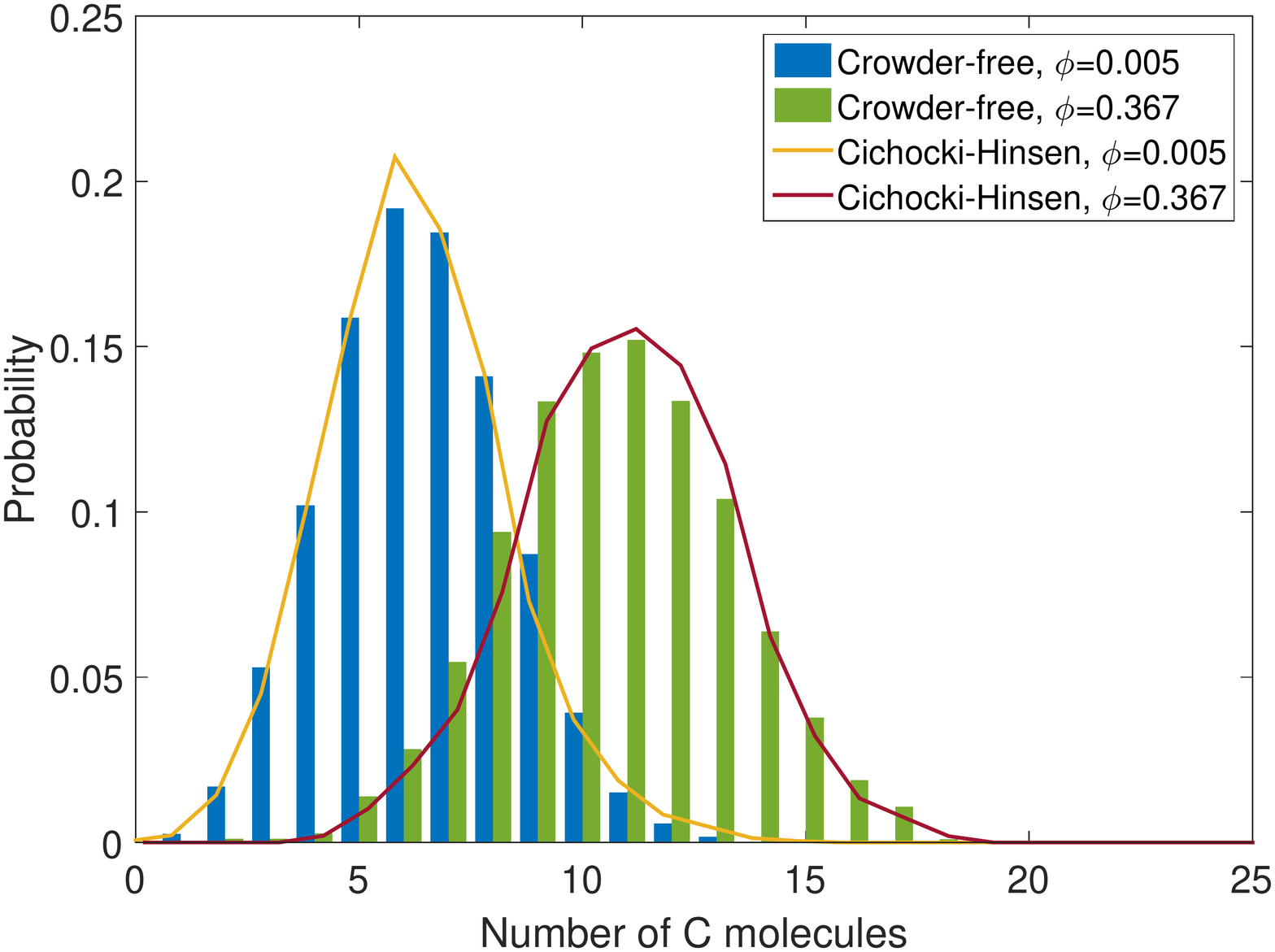}
\caption{Equilibrium distribution of the number of $C$ molecules for the reaction $A+B \xrightleftharpoons[]{}C$. Each distribution is a time average over single long trajectory of length $10^5$ iterations. Parameter values are $V=1$, $R=0.05$, $\Delta t=10^{-4}$, $D_0=0.1$ for the point particle, $D_0=0.01$ for the crowders, reaction radius $r=0.025$, forward reaction rate $\lambda_1=9\times 10^3$, backward reaction rate $\lambda_2=1$, unbinding distance $\sigma=0.025$.}\label{fig3}
\end{figure}
For each algorithm, we simulated two long trajectories of a system initially consisting of $30$ uniformly distributed $A$ molecules and 30 uniformly distributed $B$ molecules, in a sea of 10 (low crowding) and 700 (high crowding) crowders. The simulation time was much longer than the time for the system to reach equilibrium. In Fig. \ref{fig3} we show the equilibrium distribution for the number of $C$ molecules. The mean number of $C$ molecules shifts from around $6$ with low crowding to around $11$ with high crowding. The crowder-free algorithm agrees almost perfectly with the Cichocki-Hinsen algorithm for both examples, thus confirming that the crowder-free algorithm accurately imitates the Cichocki-Hinsen algorithm, but with a dramatic reduction in computation time.
\subsection{A note on more complex systems}
The crowder-free algorithm proposed above specifically concerns a uniform distribution of crowders with the same radius, however the results can equally be applied to more complex systems. 

For sets of crowders with different radii, say $N_C^{(i)}$ crowders of radius $R_i$ for $i=1,...,k$, we can simply use the formula:
\begin{equation}
P(\text{illegal})=\sum_{i=1}^k\frac{N_C^{(i)}\pi R_i^2 \delta x}{V},
\end{equation}
which will give the probability of a move $\delta x$ resulting in a collision. Of course, this formula relies on the assumption that $\delta x \ll R_i$ for all $i=1,...,k$. 

For systems with a non-uniform distribution of crowders of radius $R$, the algorithm can still be used if the crowder distribution is locally uniform. In that case, we can divide the volume up into $k$ subvolumes $V_i$ with $N_C^{(i)}$ crowders for $i=1,...,k$, whre $V_1+...+V_k=V$ and $N_C^{(1)}+...+N_C^{(k)}=N_C$. Then we can apply the formula:
\begin{equation}
P(\text{illegal})=\frac{N_C^{(i)}\pi R^2 \delta x}{V_i},
\end{equation}
for a point particle in the $i^\text{th}$ subvolume. However, this method will only really work if the crowder distribution remains roughly constant in time. If the crowders are diffusing fast enough that the overall distribution flattens on the timescale of the simulation, then subvolume $i$ will not always contain $N_C^{(i)}$ crowders. Since we do not know how $N_C^{(i)}$ will change \emph{a priori}, we cannot really use the crowder-free algorithm for such examples.
\section{Finite-size particles in a crowded environment}\label{Finite}
Studying the behaviour of reactive point particles in the presence of crowders provides useful information about real biochemical systems in which the reactive particles are much smaller than the crowders they encounter. This is an accurate description of, for example, small proteins or amino acids diffusing in the vicinity of ribosomes or large enzymes. However, biochemical particles also encounter crowders with a similar size to themselves. In order to study these examples effectively, we must also be able to simulate reactive particles which occupy a non-zero volume. A version of the Cichocki-Hinsen algorithm for which the reactive particles occupy a non-zero volume is given below. Since reactive particles now have a physical radius, we no longer need to define a reaction distance for bimolecular reactions: particles react with a rate $\lambda_j$ if they physically intersect. This is known as partial-absorption Smoluchowski binding \cite{agbanusi2014comparison}. \\~\\
\textbf{Cichocki-Hinsen algorithm with finite-size reactive particles}
\begin{enumerate}
\item Uniformly distribute the reactive particles and the crowders in the volume, such that no particles (reactive or crowder) are intersecting each other. Let $N$ be the total number of particles, and randomly assign each particle a unique index $1,...,N$.

\item Uniformly sample an integer $i$ from $1,...,N$. Propose a new position for particle $i$ at a random Normal$(0,\sqrt{2 D_i \Delta t})$ displacement in each spatial dimension, where $D_i$ is the diffusion coefficient of particle $i$ and $\Delta t$ is the simulation time step. If particle $i$ is a crowder, check if this new position causes an intersection between any particles. If so, place particle $i$ back in its original position, if not, place particle $i$ in the new position. Otherwise if particle $i$ is a reactive particle, check if this new position causes an intersection between $i$ and exactly one other reactive particle and no crowders. If so, and if that particle can react with $i$, proceed to (3). Otherwise, if the new position causes any other type of intersection, place the particle back in its original position, if not, place the particle in its new position. Proceed to (4).

\item Propose a bimolecular reaction $j$ with probability $\lambda_j \Delta t$, where $\lambda_j$ is the corresponding reaction rate. If successful, check if any daughter particles would intersect another particle. If so, skip the reaction, place particle $i$ back in its original position; if not, allow the reaction to proceed. 

\item For each reactive particle of a type involved in a unimolecular reaction $j$, propose a reaction with probability $\lambda_j \Delta t/N$, where $\lambda_j$ is the reaction rate. If successful, check if any daughter particles would intersect any other particles. If so, skip the reaction; if not, allow the reaction to proceed.

\item For each zero-order reaction $j$, propose a reaction with probability $\lambda_j \Delta t/N$, where $\lambda_j$ is the reaction rate. If successful, check if any of the new particles would intersect another particle. If so, skip the reaction; if not, allow the reaction to proceed.

\item Advance time by $\Delta t/N$. Let $N$ be the new total number of particles and randomly reassign each particle a unique index $1,...,N$. Return to (2) and repeat until a target time has elapsed.
\end{enumerate}

Note that this algorithm is distinct from the Cichocki-Hinsen algorithm in Section \ref{PointParticles} in several ways, mainly because in this algorithm time is advanced by $\frac{\Delta t}{N}$ at each time step. This is because here step (3) is nested inside step (2). The reason for this is that bimolecular reactions occur in this algorithm when two reactive particles physically intersect. This is an illegal move, and if the particles do not react then they must not be allowed to remain in that position, but rather revert to the previous position, hence bimolecular reactions and diffusion are closely coupled in this algorithm. It follows that $N$ can change during steps (2)-(3), and so it does not make sense to place step (2) inside a for-loop over $i=1,...,N$.

Again, step (2) is the overwhelmingly time consuming step for this algorithm, so as before we will attempt to find an expression giving the probability that a given jump causes an intersection with a crowder. However, we will not be able to get substantial speed gains on the same scale that we obtained with point-particles, because now even a crowder-free algorithm will contain finite-size reactive particles. Our speed increase will arise from removing a subset of the volume-occupying particles (the crowders) rather than all of them, as before. Obviously, our method will work best if there are many more crowders than reactive particles, though it will always be faster than the standard algorithm. 
\subsection{Derivation}
To derive an analogous formula to Eq. \eqref{pillegal} for the finite-volume case, consider a reactive particle with radius $r>0$ attempting to move a distance $\delta x$ in a sea of $N_C$ uniformly distributed crowders of radius $R$. In Section \ref{PointParticles}, we observed that, to first order in $\frac{\delta x}{R}$, the probaiblity of a reactive particle performing an illegal move depends only on its behaviour in the vicinity of a single crowder. However, a particle of radius $r$ moving near a single crowder of radius $R$ is identical to a point-particle moving near a crowder of radius $R+r$: in both cases, the two particle centres are forbidden from being nearer than $R+r$ from each other. It follows that Eq. \eqref{pille} can be easily adapted for use in this section, but with $R$ replaced by $R+r$. In other words, we can simply write:
\begin{equation}
P(\text{illegal})=\frac{\pi N_C (R+r)^2 \delta x}{V}+o\left(\frac{\delta x}{R+r}\right).
\end{equation}
Observe that we do not need to consider the probability of intersecting reactive particles here. This is because the reactive particles will all be simulated explicitly, so a collision between reactive particles in the crowder-free algorithm will be simulated identically to the original algorithm.

As before, we will also need to moderately adapt the reaction part of our algorithm. Again, if a daughter particle is created a small distance $\sigma$ from a parent particle, and $\sigma \ll R+r$, then we can use the formula $P(\text{illegal})=\frac{\pi N_C (R+r)^2 \sigma}{V}$. Note, however, that this is much less likely to occur with finite-size particles, since $\sigma$ will typically be a similar order of magnitude to $r$, which is in turn typically a similar order of magnitude to $R$. Exceptions include the monomolecular conversion reaction $A \rightarrow B$, but even this may pose problems if the radius of $B$ is larger than that of $A$. For almost all reactions we therefore use the probability that a uniformly distributed point in space can accomodate a particle of radius $r$. 

This probability is not the simple expression used in Section \ref{PointParticles}, rather it derives from scaled particle theory (SPT). The reason for this is that there are unoccupied points in space which are inaccessible to the particle of radius $r$. These are the points which do not lie inside a crowder but do lie within a distance $R+r$ from a crowder's centre. SPT has been used to obtain analytical expressions for the effect of crowding on intrinsic noise in two-dimensional systems, and was observed to give very accurate results \cite{grima2010intrinsic}. In three dimensions, it offers an expression for the probability that a uniformly distributed point in space of volume $V$ can accomodate a particle of radius $r$, given that the space contains $N_C$ crowders of radius $R$ \cite{zimmerman1991estimation}:
\begin{align}\label{SPT}
\text{log}&\left[P(\text{legal})\right]=\text{log}(1-\phi)-\frac{Br}{1-\phi}-\frac{4\pi Ar^2}{1-\phi}-\frac{B^2r^2}{2(1-\phi)^2}\nonumber\\
&-\frac{4\pi}{3} \left[ \frac{N_C}{V(1-\phi)}+\frac{B^2C}{3(1-\phi)^3}+\frac{AB}{(1-\phi)^2}\right]r^3,
\end{align}
where $A=\frac{N_CR}{V}$, $B=\frac{4 \pi N_C R^2}{V}$, and $C=\frac{N_CR^2}{V}$. The crowder-free algorithm for finite-size reactive particles is then as follows:\\~\\
\textbf{Crowder-free algorithm with finite-size reactive particles}
\begin{enumerate}
\item Uniformly distribute the reactive particles in the volume, such that no particles are intersecting each other. Let $N$ be the total number of particles, and randomly assign each particle a unique index $1,...,N$.

\item Uniformly sample an integer $i$ from $1,...,N$. Propose a new position for particle $i$ at a random Normal$(0,\sqrt{2 D_i \Delta t})$ displacement in each spatial dimension, where $D_i$ is the diffusion coefficient of particle $i$ and $\Delta t$ is the simulation time step. With probability $\frac{\pi N_C (R+r)^2 \delta x}{V}$, where $r$ is the radius of particle $i$, put the particle back in its original position. Otherwise, check if this new position causes an intersection between $i$ and exactly one other particle. If so, and if that particle can react with $i$, proceed to (3). Otherwise, if the new position causes any other type of intersection, place the particle back in its original position, if not, place the particle in its new position. Proceed to (4).

\item Propose a bimolecular reaction $j$ with probability $\lambda_j \Delta t$, where $\lambda_j$ is the corresponding reaction rate. If successful, evaluate $P(\text{legal})$ according to Eq. \eqref{SPT} for each daughter particle. Let $p$ be the product of each $P(\text{legal})$. With probability $1-p$, skip the reaction, place particle $i$ back in its original position. Otherwise check if any daughter particles would intersect another particle. If so, skip the reaction, place particle $i$ back in its original position; if not, allow the reaction to proceed.

\item For each reactive particle of a type involved in a unimolecular reaction $j$, propose a reaction with probability $\lambda_j \Delta t/N$, where $\lambda_j$ is the reaction rate. If the reaction is of the type $A \xrightarrow{} B$ and the radius of $B$ is less than or equal to that of $A$, allow the reaction to proceed. Otherwise, evaluate $P(\text{legal})$ according to Eq. \eqref{SPT} for each daughter particle. Let $p$ be the product of each $P(\text{legal})$. With probability $1-p$, skip the reaction. Otherwise, check if any daughter particles would intersect any other particles. If so, skip the reaction; if not, allow the reaction to proceed.

\item For each zero-order reaction $j$, propose a reaction with probability $\lambda_j \Delta t/N$, where $\lambda_j$ is the reaction rate. If successful, evaluate $P(\text{legal})$ according to Eq. \eqref{SPT}. With probability $1-P(\text{legal})$, skip the reaction. Otherwise check if any of the new particles would intersect another particle. If so, skip the reaction; if not, allow the reaction to proceed.

\item Advance time by $\Delta t/N$. Let $N$ be the new total number of particles and randomly reassign each particle a unique index $1,...,N$. Return to (2) and repeat until a target time has elapsed.
\end{enumerate}

There is one significant case for which our crowder-free algorithm will not give accurate results, namely if the crowders are stationary and the level of crowding is high. Simulating such systems with Cichocki-Hinsen reveals that reactive particles can get trapped in regions surrounded by stationary crowders, and simply stay there for the entirety of the simulation without reacting or moving significantly. Obviously, these cases cannot be covered by the crowder-free algorithm because all reactive particles (of the same radius) have the same probability of diffusing at any time. We therefore recommend not using the crowder-free algorithm for systems with stationary crowders unless the level of crowding is sufficiently low that no trapping regions could exist. Note that this is not a problem if the reactive particles are point-particles, because they occupy no volume and will always be able to escape from a trapping region.
\subsection{Comparative tests}
\begin{figure}[h]
\includegraphics[scale=0.3]{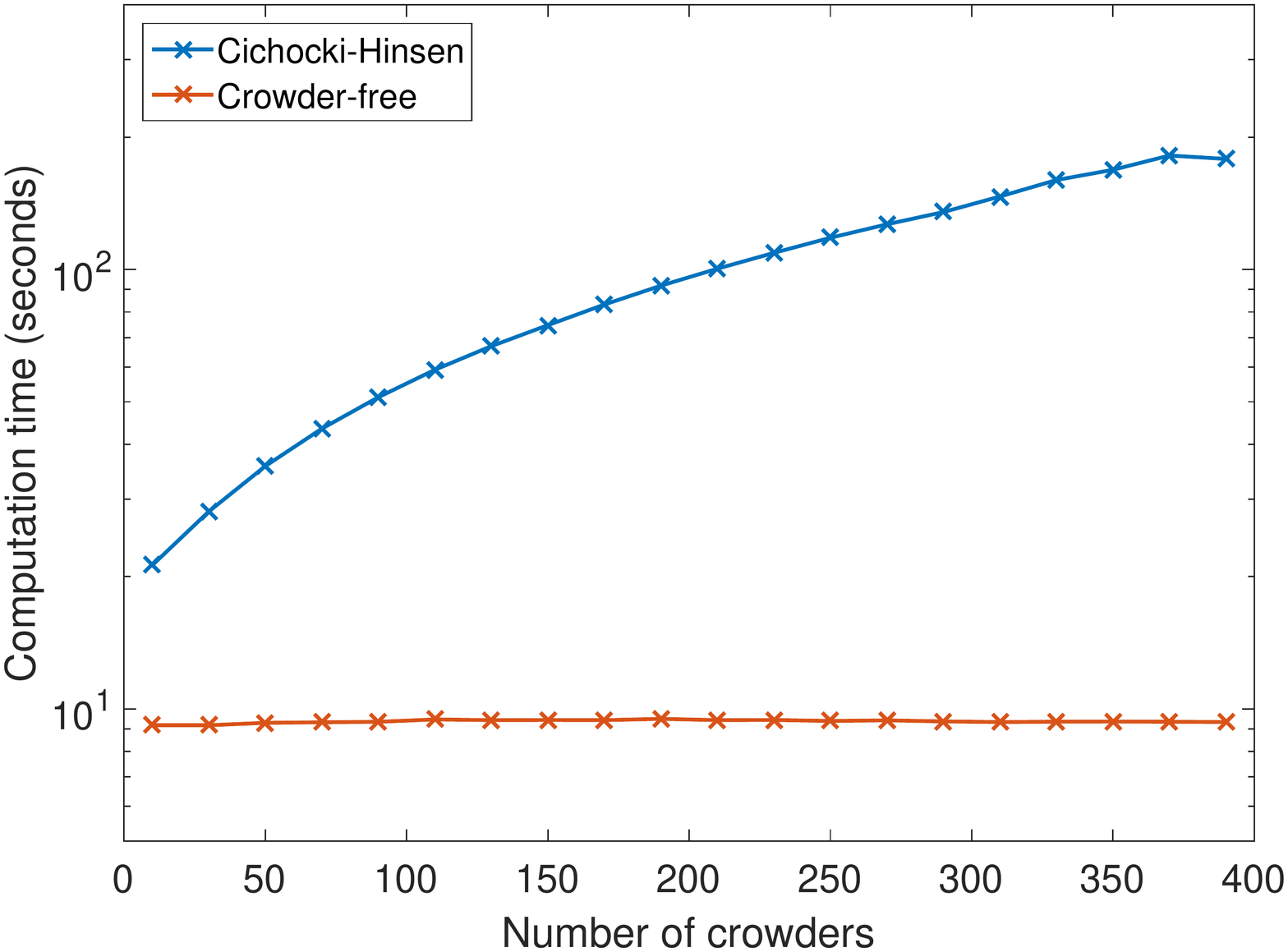}
\caption{Time taken for 100 time steps of both the Cichocki-Hinsen algorithm (blue) and the crowder-free algorithm (red), for 50 finite-size particles diffusing in crowded space. With only 10 crowders, the crowder-free algorithm is more than twice as fast. With 400 crowders, the crowder-free algorithm is over $20$ times faster. Parameter values are $V=1$, $R=0.05$, $r=0.02$, $\Delta t=10^{-5}$, $D=0.1$ for the point particle, $D=0.1$ for the crowders.}\label{fig4}
\end{figure}
In this section we perform similar tests on the crowder-free algorithm for finite-size particles to those we performed in section \ref{pptest}. We initially test the time taken for both methods to simulate pure diffusion in the presence of an increasing number of crowders. To ensure that the results are different from those in section \ref{pptest}, we now simulate 50 diffusing ``reactive'' particles (so-called even though they do not react in this example) in a sea of crowders. Of course, we do not expect to get anywhere near the 1000-fold speed increase that we achieved for the point-particle case: even with no crowders, we have to simulate 50 volume-occupying molecules, constantly ensuring that they do not intersect. 

The results of this test are plotted in Fig. \ref{fig4}. With 10 crowders, the crowder-free algorithm takes half the time of the Cichocki-Hinsen algorithm, while with 400 crowders, the crowder-free algorithm has a speed increase of over 20 times. Even for finite-size particles, therefore, the crowder-free algorithm offers a considerable speed increase, and its lack of dependence on crowder number makes it especially useful for studying high levels of crowding. 
\begin{figure}[h]
\includegraphics[scale=0.3]{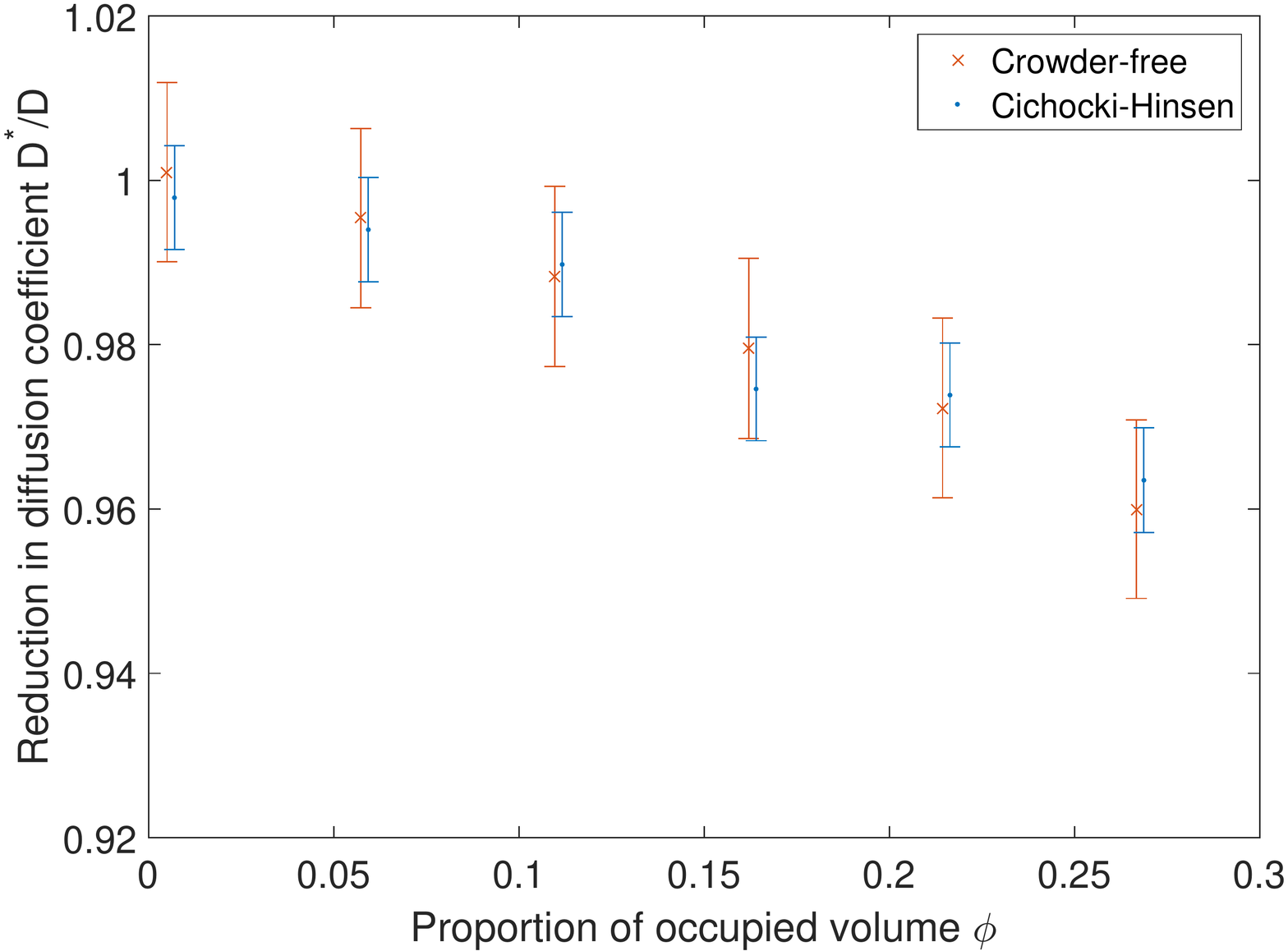}
\caption{Relative reduction in short-time diffusion coefficient for both the Cichocki-Hinsen algorithm (blue) and the crowder-free algorithm (red), for a single point particle diffusing in space, as a function of the proportion of occupied volume $\phi$. All data points are an average of 20 particles from a single simulation, error bars are 1 standard deviation. Parameter values are $V=1$, $R=0.05$, $r=0.02$, $\Delta t=10^{-5}$, $D=0.1$ for the point particle, $D=0.1$ for the crowders.}\label{fig5}
\end{figure}

The next test we perform compares estimates of short-time diffusion coefficients from the two algorithms. In both cases, we simulate $20$ finite-size particles diffusing in a sea of crowders. Because of this, a single simulation gives 20 different estimates of the diffusion coefficient. In Fig. \ref{fig5} we plot the mean (points) and standard deviation (error bars) of this sample of 20, for a variety of levels of crowding. Since the ``reactive'' particles themselves occupy a volume, we incorporate this into our calculation of the proportion of occupied volume $\phi$. As in Fig. \ref{fig2}, the two algorithms agree, with errorbars intersecting for each data point. Note that, compared to Fig. \ref{fig2}, the diffusion coefficient is reduced more for the same level of crowding. This confirms the intuitive hypothesis that finite-size particles are more influenced by crowding than point particles.
\begin{figure}
\includegraphics[scale=0.3]{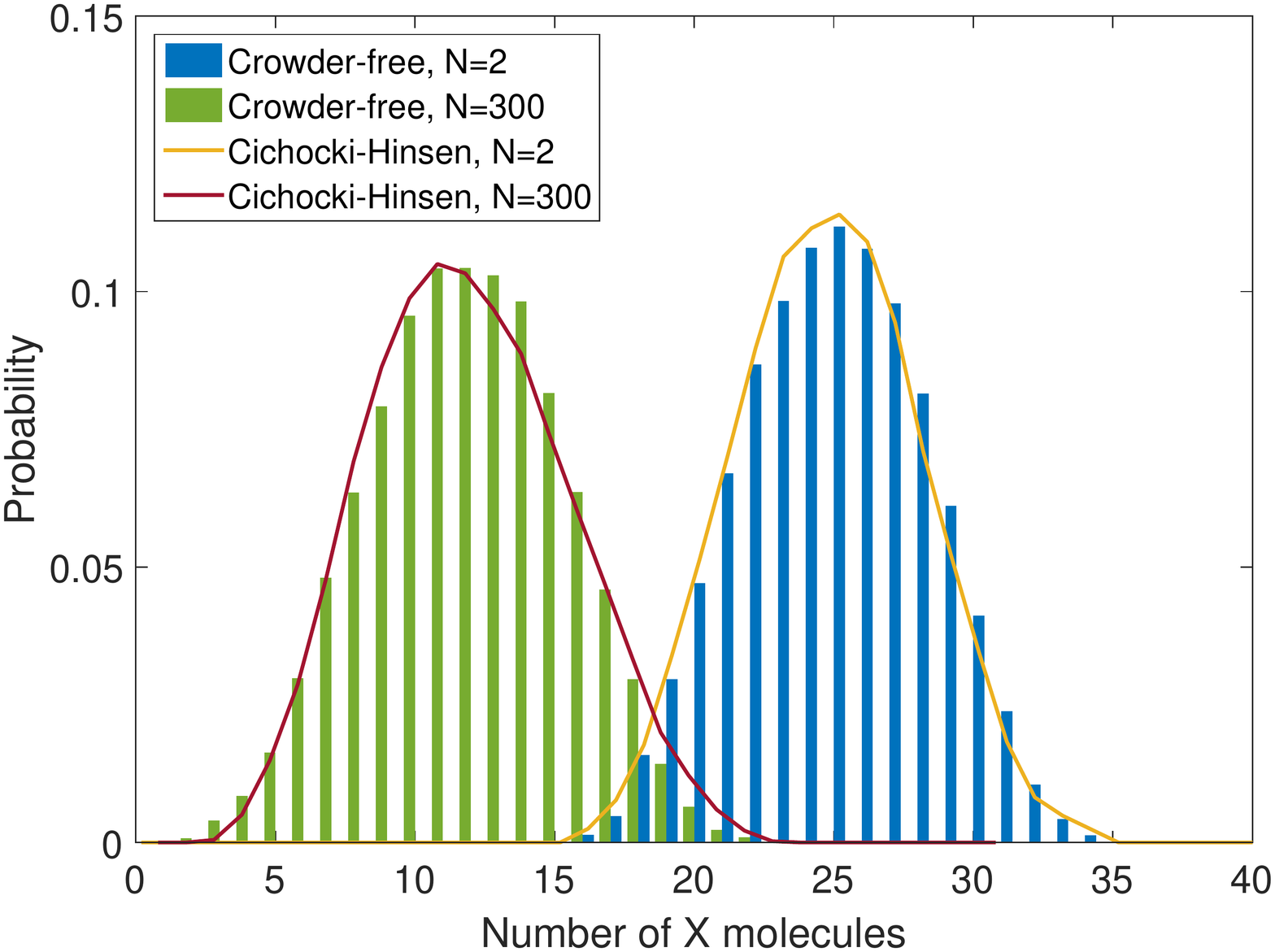}
\caption{Equilibrium distributions of the reaction $\emptyset \xrightarrow{} X,~X+X \xrightarrow{} \emptyset$ for both crowder-free algorithm (histograms) and Cichocki-Hinsen algorithm (lines) for low (blue, yellow) and high (green, red) crowding conditions. Each distribution is a time average over single long trajectory of length $10^5-10^7$ iterations. The crowder-free algorithm generally requires many fewer iterations than Cichocki-Hinsen, because the total number of particles is lower. Parameter values are $V=1$, $R=0.05$, $r=0.05$, $\Delta t=3 \times 10^{-5}$, $D=0.1$ for the point particle, $D=0.1$ for the crowders. For the forward reaction, $\lambda_1=2 \times 10^2$, for the backward reaction, $\lambda_2=3 \times 10^4$.}\label{fig6}
\end{figure}

Finally, we compare the algorithms' performance at estimating an equilibrium distribution of a chemical reaction. This time we simulate the reaction $\emptyset \xrightarrow{} X,~X+X \xrightarrow{} \emptyset$, in which particles are created at uniformly distributed points in space and react with a fixed rate when they collide. This system has previously been studied spatially as an example of protein synthesis and degradation \cite{smith2016analytical}. We expect that, contrary to the example in Fig. \ref{fig3}, crowding will reduce the mean number of $X$, since the creation of $X$ will be less likely in crowded conditions. 

In Fig. \ref{fig6} we plot the equilbrium distribution of the number of $X$ molecules for both algorithms in both low and high crowding conditions. Each distribution is calculated as a time average over a single long trajectory, of between $10^5$ and $10^7$ iterations. The crowder-free algorithm clearly requires fewer iterations than Cichocki-Hinsen because each iteration of both algorithms advances time by $\frac{\Delta t}{N}$ where $N$ is the total number of particles, and Cichocki-Hinsen generally has many more particles to simulate. As predicted, the mean of the distribution is much lower in the high crowding example than the low crowding example. As with all previous tests, the crowder-free algorithm agrees almost perfectly with the Cichocki-Hinsen algorithm, confirming that our algorithm suffers little apparent loss of accuracy compared to the Cichocki-Hinsen algorithm, despite its considerable speed increases. Note that we do not calculate $\phi$ for these examples, because the number of reactive particles fluctuates over time, and therefore so does $\phi$.
\section{Discussion}\label{Discussion}
In this paper, we have proposed a modification to the commonly used Cichocki-Hinsen Brownian dynamics algorithm for simulating reaction-diffusion systems in a crowded environment. We call our modified algorithm a \emph{crowder-free} algorithm because we don't simulate crowders explicitly. Instead, we rigorously derive the probability that a small displacement of size $\delta x$ would result in a collision with a crowder. This implies that, instead of simulating crowders, we can simply reject each attempted particle displacement with precisely that probability. 

We tested our algorithm in terms of both speed and accuracy, both for cases with reactive point particles and with finite-size reactive particles. The crowder-free algorithm always provides a speed increase over the underlying Cichocki-Hinsen algorithm: this speed increase varied from $2$ to over $1000$ for the set of examples studied in this paper. Furthermore, the crowder-free algorithm provides data which is near-indistinguishable from the data extracted from the corresponding Cichocki-Hinsen algorithm: this was shown to be true for both diffusive and reactive information. The crowder-free algorithm therefore shows no apparent loss of accuracy compared to the Cichocki-Hinsen algorithm, which, coupled with the clear speed increases, makes it a very attractive algorithm for simulating chemical reactions in a crowded environment. 

There are two main cases where  the crowder-free algorithm is not more effective than the Cichocki-Hinsen algorithm. Firstly, if the initial crowder distribution is not uniform and spreads out over time. In that case, our algorithm is inadequate because we do not know \emph{a priori} how fast the crowders will diffuse. Note, however, that non-uniform crowder distributions are not a problem in themselves: we can simply subdivide the volume into regions where the distribution is locally uniform, and derive separate values of $P(\text{illegal})$ in each region. The second case involves stationary crowders. If the level of crowding is high and the crowders do not diffuse, then some regions of space may be entirely segregated from others. Since the crowder-free algorithm allows all reactive particles to diffuse anywhere in space, it cannot accurately imitate the Cichocki-Hinsen algorithm in this case.

Finally, we note that further speed increases in both the crowder-free algorithm and the Cichocki-Hinsen algorithm may be possible by more efficient methods of measuring the distance between particles. One smart idea, used in Ref. \cite{andrews2004stochastic}, is to subdivide the volume into regions and only check distances between particles in the same or neighbouring regions. We did not use such methods in this paper so as to not overcomplicate the algorithms, however any implementations of our algorithm would certainly benefit from these techniques. 

\acknowledgments
This work was supported by a BBSRC EASTBIO PhD studentship to S.S. and by a Leverhulme grant award to R.G. (RPG-2013-171).

\bibliography{mybibfile}{}
\end{document}